# Resilient Neural-Variable-Structure Consensus Control for Nonlinear MASs with Singular Input Gain Under DoS Attacks

Ladan Khoshnevisan[*], Xinzhi Liu

*Abstract*— This paper proposes a reliable learning-based adaptive control framework for nonlinear multi-agent systems (MASs) subject to Denial-of-Service (DoS) attacks and singular control gains, two critical challenges in cyber-physical systems. A neural-variable-structure adaptive controller is developed to achieve leader-follower consensus while ensuring robustness to external disturbances and adaptability to unknown nonlinear dynamics. A reliability-assessment rule is introduced to detect communication loss during DoS attacks, upon which a switched control mechanism is activated to preserve closed-loop stability and performance. Unlike existing resilient MAS control methods, the proposed strategy explicitly accommodates *singular* control gains and does not rely on restrictive assumptions such as Lipschitz continuity or prior bounds on nonlinearities. To the authors' knowledge, this is the first work to integrate neural learning, variable-structure robustness, and reliability-based switching into a unified consensus-tracking control architecture for heterogeneous nonlinear MASs with singular input gains under DoS attacks. Lyapunov-based analysis establishes uniform ultimate boundedness of all closed-loop signals, and Matlab/Simulink simulations on a connected automated vehicle platoon demonstrate the method's effectiveness and resilience.

## 1 INTRODUCTION

RELIABLE control of multiagent systems (MASs) have garnered significant interests of researchers from different aspects, such as cyber physical systems [1], intelligent transportation systems [2], [3], and connected automated vehicles [4]. In spite of advantages of distributed control and networking configuration, like scalability and flexibility in large-scale MASs, the system is exposed to several cyber security issues [5], [6], such as deception attacks, spoofing and message falsification [7], [8], [9], and Denial of service (DoS) [10], [11], [12]. Therefore, it is essential to propose a resilient procedure that can deal with the distributed consensus control of MASs.

DoS attack is one of the most important cyber-attacks, as the communication link between the system components will be engaged by the adversaries, which means that no controller signal can be transferred to the subjected agent [13], [14]. Therefore, it can lead to instability and dangerous happenings in the system. Most of the proposed methods, dealing with DoS attacks in MASs, are designed based on the linearized model [9], [11], [15], [16], which ignores important specifications of the nonlinear part of the system description. On the other hand, the proposed procedures which are designed for nonlinear MASs are confined by constraining limitations, like knowledge of the nonlinear part [17], [18] or considering Lipschitz condition for unknown nonlinearity [9], [19], [20], [21], which drastically limits their applicability on real systems.

Some authors have proposed nonlinear resilient procedures for MAS applications based on attack detection through which the subjected agent will be omitted from the network topology and no control is provided for the corresponding agent, which can result in being unstable [7], [22], [23]. In addition, a secure control procedure is proposed for nonlinear networked control systems in [24] where the inverse function of control input gain is implemented in the control approach, and an observer-based distributed control approach is designed in for nonlinear MASs with nonsingular control gain. As it is impractical to always ensure that the control input gain is nonsingular, the proposed approach is impractical to the systems with singular control gain.

In addition to the above limitations, recent studies have attempted to enhance the resilience of MASs under DoS attacks through event-based and switching communication frameworks. For example, the event-triggered neural resilient control framework in [25] addresses switching communication topologies and mixed connectivity-maintained/broken DoS attacks using multiple Lyapunov functions and dynamic triggering mechanisms. However, these methods rely on strictly nonsingular input gains and require structural assumptions on subsystem switching and network connectivity, making them unsuitable for systems where the control gain may become singular, a common characteristic in many cyber-physical and robotic applications. Moreover, although adaptive neural estimation is employed, the design remains tied to event-triggered schemes and does not accommodate continuous reliability assessment or controller switching during communication failures.

Similarly, the event-driven formation framework proposed in [26] investigates partial component consensus using robust event-based and self-triggered strategies. While effective for reducing

[*]L. Khoshnevisan
Department of Mechanical and Mechatronics Engineering, University of Waterloo, Waterloo, ON, Canada, N2L 3G1

X. Liu
Department of Applied Mathematics, University of Waterloo, Waterloo, ON, Canada, N2L 3G1

E-mail addresses: lkhoshnevisan@uwaterloo.ca (L. Khoshnevisan) xzliu@uwaterloo.ca (X. Liu)

communication frequency and achieving partial synchronization, these works do not address unknown nonlinear dynamics through adaptive learning, nor do they consider reliability-based switching control or singular control gains. Additionally, the focus on partial component behavior and fixed CAV dynamics differs fundamentally from the general heterogeneous nonlinear MAS setting considered in this paper.

Consequently, despite substantial progress, no existing approach simultaneously addresses: (i) nonlinear heterogeneous MASs with unknown and unrestricted nonlinearities (without Lipschitz assumptions), (ii) singular control gains, (iii) DoS-induced controller signal loss, and (iv) a reliable adaptive control mechanism that integrates neural learning, variable-structure robustness, and a switched architecture activated during communication failures. Motivated by these gaps, this paper develops a reliable adaptive neural–variable-structure control framework capable of guaranteeing consensus tracking and uniform ultimate boundedness for MASs despite DoS attacks, unknown nonlinearities, and inherent singularities in the control gain. The key contributions are summarized as follows:

1) **Novel reliable adaptive framework for nonlinear heterogeneous MASs with *singular control gain* under DoS attacks:**

A neural–variable-structure adaptive controller is developed that achieves reliable leader–follower consensus despite *singular or rank-deficient* control gains and intermittent DoS attacks. To the authors' knowledge, this is the first MAS control framework that simultaneously addresses DoS-induced input loss, nonlinear unknown dynamics, and singular control gains without requiring Lipschitz conditions, known dynamics, or gain invertibility.

2) **Rigorous reliability rule and switched control design:**

A mathematically explicit reliability assessment rule is integrated with the control strategy, providing an online mechanism to detect the loss of communication or performance degradation due to DoS. A switched control mechanism is then triggered to maintain system stability and performance, significantly enhancing resilience against network-based attacks.

3) **Adaptive neural estimation of unknown nonlinearities without restrictive structural assumptions:**

The controller incorporates an online neural estimator that approximates the unknown nonlinear dynamics in real time, eliminating the need for known bounds, Lipschitz assumptions, or prior modeling. This significantly expands applicability to real cyber-physical systems with complex, uncertain, or irregular nonlinearities.

4) **Unified stability and robustness guarantees under singularity, uncertainty, and DoS:**

A comprehensive stability analysis is provided, establishing uniform ultimate boundedness (UUB) for all closed-loop signals even in the presence of external disturbances, uncertain parameters, singular input gain matrices, and intermittent DoS attacks. The analysis integrates the neural estimator, variable-structure dynamics, and switched control mechanism into a single coherent stability framework.

5) **Demonstration of resilience and performance benefits in realistic cyber-physical scenarios:**

The effectiveness and reliability of the proposed approach are verified through extensive Matlab/Simulink simulations on a connected automated vehicle (CAV) platoon, demonstrating improved resilience, reduced tracking errors, and robustness to both singularities and DoS attacks compared to existing methods.

The remainder of the paper is organized as follows: The problem formulation of a MAS under DoS attacks is discussed in Section 2. Then, designing and stability analysis of a secure artificial-intelligence-based adaptive resilient control procedure are provided in Section 3. After that, the proposed procedure is evaluated via simulation results of a numerical example in Section 4. Finally, some conclusion remarks are provided through Section 6.

## 2 PROBLEM FORMULATION: MAS UNDER DOS ATTACKS

We consider a multi-agent system of $N$ heterogeneous following agents and a leader. The agents are connected through a nonreliable network, which can be affected by denial of service (DoS) attacks. In other words, cyber-attacks in the network layer, through which an adversary tries to engage the communication link and disrupts the connection, are considered in this paper. Furthermore, a general network topology, which will be described later, is assumed to make the proposed procedure compatible to different types of connections. In what follows, the network topology through which the agents are connected to each other, and modelling of a heterogeneous multi-agent system and DoS attacks are provided. As the controller signal cannot be updated properly during the DoS attacks, the corresponding agent to be controlled will be out-of-control or in the lack of required information. So, it is imperative to design a procedure, which can be effective even during the DoS attack intervals.

### 2.1 Network communication topology

The communication network topology of a multi-agent system is considered to be a directed graph of wireless links, containing a spanning tree with the leader as the root, through which each node represents an agent. Therefore, a MAS including $N$ followers can be modeled by a $(N+1)$-order directed graph $\aleph(\Upsilon, \Omega, A)$, where $\Upsilon = (0,1,\dots,N)$ is the set of nodes in which 0 indicates the leader, $\Omega = \Upsilon \times \Upsilon$ is the set of directed edges, and $A = [\alpha_{ij}]_{(N+1)\times(N+1)}$ is the adjacency matrix, in which $\alpha_{ij} = 1$, if and only if node $i$ can receive information from node $j$, but not necessarily vice versa, otherwise $\alpha_{ij} = 0$. In other words, elements of the adjacency matrix indicate the directed connection between each two agents. Moreover, as the leader can only send information to other agents, but does not receive data from followers, we have $\alpha_{0j} = 0$ ; $\forall j = 0,\dots,N$.

### 2.2 Heterogeneous MAS modeling

In this part, a multi-agent system with $N$ heterogeneous follower and a leader is considered. The information of the agents, including the followers and the leader, as well as the control input vector for each follower, provided by the corresponding controller, are transferred via a nonreliable V2V network structure. The system description of the $i^{th}$ following agent is considered as:

$$\begin{aligned}\dot{x}_{1i}(t) &= x_{2i}(t), \\ \dot{x}_{2i}(t) &= x_{3i}(t), \\ &\vdots \\ \dot{x}_{ni}(t) &= a_i x_{ni}(t) + b_i u_i(t) \\ &\quad + f_i\big(x_{1i}(t), x_{2i}(t), \ldots, x_{ni}(t)\big) \\ &\quad + \omega_i(t),\end{aligned} \quad (1)$$

in which $x_{1i}(t)$, $x_{2i}(t)$, ..., $x_{ni}(t)$ are the states of the $i^{th}$ following agent, $u_i(t)$ is the control input signal, which will be designed in the next section, $f_i(x_{1i}(t), x_{2i}(t), \ldots, x_{ni}(t))$ is the unknown nonlinear part of the system description presented as a nonlinear function in $\mathbb{R}^n \to \mathbb{R}$, and $\omega_i(t)$ stands for the external disturbances, which assumed to be bounded as follows:

$$\|\omega_i\| \leq \omega_i^*, \quad (2)$$

where $\omega_i^*$ is a known positive constant. Now, we suppose $X_i(t) = [x_{1i}(t) \cdots x_{ni}(t)]$. Therefore, system description (1) can be rewritten as the following concise form:

$$\begin{aligned}\dot{X}_i(t) &= A_i X_i(t) + B_i u_i(t) + B_{fi} f_i(X_i(t)) + B_{\omega i} \omega_i(t), \\ Y_i(t) &= C_i X_i(t),\end{aligned} \quad (3)$$

where $X_i(t)$ is the state vector, $u_i(t)$ is the control input signal. Furthermore, the leader is supposed to be modeled by the following known system description:

$$\begin{aligned}\dot{X}_0(t) &= A_0 X_0(t) + B_{f0} f_0(X_0(t)), \\ Y_0(t) &= C_0 X_0(t).\end{aligned} \quad (4)$$

Furthermore, $A_i$, $B_i$, $B_{fi}$, $B_{\omega i}$, and $C_i$, for $i = 0, \ldots, N$ are the constant matrices with appropriate dimensions as follows:

$$A_i = \begin{bmatrix} 0 & 1 & 0 & 0 \\ 0 & 0 & 1 & 0 \\ \vdots & \vdots & \vdots & \vdots \\ 0 & 0 & 0 & a_i \end{bmatrix}, B_i = \begin{bmatrix} 0 \\ 0 \\ \vdots \\ b_i \end{bmatrix}, B_{fi} = B_{\omega i} = \begin{bmatrix} 0 \\ 0 \\ \vdots \\ 1 \end{bmatrix}, C_i = I_n. \quad (5)$$

**Remark 1:** The leader agent is assumed to be a command generator exosystem [27], that generates the information about the reference behavior and can have any kind of known model, regardless of the system description of the followers. In other words, the leader is not responsible to send the control input signal to the following agents and just sends the reference data to other agents. Furthermore, the controller will be proposed to make sure that all the following agents track the desired target trajectory provided by the leader, although there exist external disturbances and DoS attacks in the system.

**Remark 2:** In contrast to the prevalent papers which have mainly considered Lipschitz condition for the unknown nonlinear part of the system, which is a drastically limiting constraint in practical applications, in this paper we design a neural network to estimate the nonlinearity of the system description.

**Remark 3:** The leader agent, which provides the reference values for the other agents, is assumed to be an autonomous system on which no adversary impacts its communication links to the other agents and controllers.

The unknown nonlinear function is modelled via the following neural network:

$$f_i(x_i) = \eta_i^T \varphi_i(X_i), \quad (6)$$

where $\eta_i = [\eta_{1i} \cdots \eta_{ni}]^T$ and $\varphi_i = [\varphi_{1i} \cdots \varphi_{ni}]^T$, denote the desired weight vector and the activation function vector of the neural network, in which $\eta_{li} = [\eta_{li}^1, \ldots, \eta_{li}^m]$ and $\varphi_{li} = [\varphi_{li}^1(x_{li}), \ldots, \varphi_{li}^m(x_{li})]$ for $l = 1, \ldots, n$, defined by (7), and $m$ represents the number of the neurons, $X_i$ is the state vector of the $i^{th}$ follower.

$$\varphi_{li}^k(x_{li}) = \tanh(x_{li}), k = 1, \ldots, m; l = 1, \ldots, n. \quad (7)$$

Furthermore, the desired weight vector $\eta_i$ is assumed to satisfy

$$\|B_{fi} \eta_i^T\| \leq \eta_i^*, \quad (8)$$

where $\eta_i^*$ is a known positive constant. As the nonlinearity in the system description is an unknown function in system description (1), the following adaptive neural network estimator is considered:

$$\hat{f}_i(X_i) = \hat{\eta}_i^T \varphi_i(X_i), \quad (9)$$

in which $\hat{\eta}_i = [\hat{\eta}_{1i} \cdots \hat{\eta}_{ni}]^T$, where $\hat{\eta}_{li} = [\hat{\eta}_{li}^1, \ldots, \hat{\eta}_{li}^m]^T$ for $l = 1, \ldots, n$ is the adaptive weight vector to be proposed later, and satisfies the following relationship

$$\|\hat{\eta}_i - \eta_i\| \leq \varepsilon_i^*, \quad (10)$$

where $\varepsilon_i^*$ is a known positive constant. For the sake of reliability assessment of the controller, the following reliable rule model is considered:

$$u_i(t) = (1 - \epsilon_i) u_{ia}(t) + \epsilon_i u_{ia}(k), \quad (11)$$

where $\epsilon_i$ is a scalar defined to be 0 in the case of no DoS attack in the system and the controller doesn't fail. Furthermore, $\epsilon_i = 1$ when the DoS attack is activated, and the controller fails to be transmitted. In addition, $u_{ia}(t)$ is the controller input signal in the absence of attacks. More explanation is given in the following Sections. Furthermore, we define the set $\{t_k\}_{k \in N}$ as the sequence of time instants that the closed loop system begins out of control for the $k^{th}$ time, and $u_{ia}(k)$ is the value of control input $u_{ia}(t)$ at time instant $t_k$.

### 2.3 Denial of Service Modelling

In the case of DoS attack, the subjected communication link is blocked. Regardless of the link that is subjected by the adversaries, the frequency and the energy of intermittent DoS attacks are assumed to be constrained. Let $\Lambda(0, \infty)$ denotes the union of time intervals in which no DoS attack is present in the system and communication is allowed, and $\Xi(0,\infty)$ is the union of time intervals when the communication between a controller and the corresponding agent is denied.

**Remark 4:** The reliability rule is triggered whenever no new packet or state update is received within an expected time window. This is consistent with standard DoS detection mechanisms used in networked control.

**Assumption 1:** (Constraint on DoS frequency) [13] The maximum number of DoS attacks occurred during interval $[\tau, t)$, denoted by $n(\tau, t)$, satisfies the following relationship:

$$n(\tau, t) \leq n_0 + \frac{t-\tau}{\tau_D}, \tag{12}$$

where $n_0 \in \mathbb{R}_{\geq 0}$ denotes the initial number of attacks at time $\tau$, and $\tau_D \in \mathbb{R}_{>0}$ presents time interval between subsequent attacks. Furthermore, $\frac{t-\tau}{\tau_D}$ represents the average number of DoS off/on transitions per unit time.

**Assumption 2:** (Constraint on DoS energy) [28] The energy of DoS attacks occurred in the communication between a controller and the corresponding agent during interval $[\tau, t)$, denoted by $\Xi(\tau, t)$, is bounded by the following relationship:

$$|\Xi(\tau, t)| \leq \frac{t-\tau}{T}, \tag{13}$$

in which $T \in \mathbb{N}_{\geq 1}$ is the maximum number of packets that can be sent by an attacker during the time interval $[\tau, t)$ in a DoS attack.

## 3 NEURAL-VARIABLE-STRUCTURE CONSENSUS CONTROL

A secure adaptive control procedure is proposed in this section, which is empowered by neural-network approach, as an artificial intelligent method, for estimating the unknown nonlinearity of a multi-agent system, where some of which may be subjected to DoS attacks. The objectives of the proposed procedure are as follows:

1) The states of the $i^{th}$ follower are tracking a desired profile for $j = 0,1, ..., N$ as follows:

$$\lim_{t \to \infty} \left(x_{1i}(t) - x_{1j}(t)\right) = (i-j)d_1,$$
$$\lim_{t \to \infty} \left(x_{2i}(t) - x_{2j}(t)\right) = (i-j)d_2, \tag{14}$$
$$\vdots$$
$$\lim_{t \to \infty} \left(x_{ni}(t) - x_{nj}(t)\right) = (i-j)d_n,$$

where $d_l$ for $l = 1, ..., n$ are constants, which means that there is a desired difference between each state of each two agents.

2) Resiliency to DoS attacks, which interrupt communication between each controller to the corresponding agent, is achieved.

3) Robustness against external disturbances such as environmental and contextual effects is obtained.

4) Adaptivity with respect to parameter changes in system description is met.

5) Compatibility with different network topologies is achieved without the requirement to redesign the procedure from one topology to another one.

### 3.1 Controller design and stability analysis in the absence of DoS

To design the control procedure, the following error signal vector is assumed:

$$E_i(t) = \begin{bmatrix} e_{1i}(t) \\ e_{12}(t) \\ \vdots \\ e_{ni}(t) \end{bmatrix} = \begin{bmatrix} x_{1i}(t) - x_{10}(t) - id_1 \\ x_{2i}(t) - x_{20}(t) - id_2 \\ \vdots \\ x_{ni}(t) - x_{n0}(t) - id_n \end{bmatrix}. \tag{15}$$

In the absence of DoS attacks in which $\epsilon_i = 0$ in (11), the adaptive artificial-intelligence-based control procedure at the level of the $i^{th}$ agent is designed as follows:

$$u_{ia}(t) = -\frac{1}{2}\sum_{\substack{j=0 \\ j \in \aleph}}^{N} \alpha_{ij} B_i^T P_i \left(E_i(t) - diag(tanh(E_i^T P_i)) * \bar{E}_j(t)\right) - \frac{1}{2}\sum_{\substack{j=0 \\ j \in \aleph}}^{N} \alpha_{ij} B_i^T P_i K_C * diag(tanh(P_i E_i(t))) * \bar{E}_j(t) + B_i^T \hat{u}_{di}(t) + B_i^T [B_i B_i^T + \varepsilon]^{-1} \bar{u}_{ia}(t), \tag{16}$$

where $\varepsilon$ is a positive definite diagonal matrix, $\bar{E}_j$ is a column vector of $e_{j_l} \tanh(\varrho e_{j_l})$ for $j \in \aleph$, and $l = 1, ..., n$, in which $\varrho > 1$ is a constant, $K_C$ is a diagonal positive definite designing matrix in which the $j^{th}$ diagonal element $K_{Cj} > 1$, and $P_i = P_i^T$ is a positive definite covariance matrix achieved by the following algebraic Riccati equation (ARE):

$$P_i A_i + A_i^T P_i - P_i \sum_{\substack{j=0 \\ j \in \aleph}}^{N} \alpha_{ij} B_i B_i^T P_i + \psi_i I_n = 0, \tag{17}$$

for a given $\psi_i > 0$, where $I_n$ is a $n \times n$ identity matrix. By substituting the control input (16) into the system description of an agent (3), and by using $B_i B_i^T [B_i B_i^T + \varepsilon]^{-1} = I - \varepsilon[B_i B_i^T + \varepsilon]^{-1}$ we have

$$\dot{X}_i(t) = A_i X_i(t) - \frac{1}{2}\sum_{\substack{j=0 \\ j \in \aleph}}^{N} \alpha_{ij} B_i B_i^T P_i \left(E_i(t) - diag(tanh(E_i^T P_i)) * \bar{E}_j(t)\right) - \frac{1}{2}\sum_{\substack{j=0 \\ j \in \aleph}}^{N} \alpha_{ij} B_i B_i^T P_i K_C * diag(tanh(P_i E_i)) * \bar{E}_j(t) + B_i B_i^T \hat{u}_{di}(t) + \bar{u}_{ia}(t) - u_{di}(t) + B_{fi} f_i(X_i(t)) + B_{\omega i} \omega_i(t), \tag{18}$$

where $u_{di} = \varepsilon[B_i B_i^T + \varepsilon I_n]^{-1} \bar{u}_{ia}$ is supposed as a virtual disturbance vector bounded by $\|u_{di}\| \leq \sigma_{di}^*$, where $\|.\|$ stands for the norm, and $\sigma_{di}^*$ is a known positive constant. Furthermore, $\bar{u}_{ia}$ is the exogeneous control vector as follows:

$$\bar{u}_{ia}(t) = -B_{fi}\hat{f}_i(X_i(t)) + B_{f0}f_0(X_0(t)) \\
\qquad - K_u \tanh(P_i E_i(t)) \\
\qquad - (A_i - A_0)X_0(t) - iA_i D_0, \tag{19}$$

where $D_0 = [d_1 \ \ldots \ d_n]^T$, and $K_u$ is the control gain matrix, which is assumed to satisfy $\|K_u\| \geq \|B_{\omega i}\|\omega_i^* + u_{di}^*$, where $u_{di}^*$ will be introduced later in (29). In addition, the following adaptive law is designed to achieve the weight vector of the nonlinear function estimator $\hat{f}_i(x_i(t))$ in (9) as an adaptive neural network.

$$\dot{\hat{\eta}}_i(t) = K_\eta^{-1} \varphi_i(x_i(t)) E_i^T P_i B_{fi}, \tag{20}$$

where $K_\eta$ is a positive definite diagonal designing matrix with appropriate dimension, which satisfies $\|K_\eta\| \leq K_\eta^*$ and $K_\eta^*$ is designed to be a small value. Furthermore, $\hat{u}_{di}$ is the estimated virtual disturbance, which is obtained by the following disturbance observer:

$$\dot{\hat{X}}_i(t) = A_i \hat{X}_i(t) - \frac{1}{2}\sum_{\substack{j=0 \\ j\in\aleph}}^{N} \alpha_{ij} B_i B_i^T P_i \left(E_i(t) - diag(\tanh(E_i^T P_i)) * E_j(t)\right) - \\
\frac{1}{2}\sum_{\substack{j=0 \\ j\in\aleph}}^{N} \alpha_{ij} B_i B_i^T P_i K_C * diag(\tanh(P_i E_i)) * \bar{E}_j(t) + \\
B_i B_i^T \hat{u}_{di}(t) + \bar{u}_{ia}(t) - \hat{u}_{di} + B_{fi}f_i(\hat{X}_i(t)), \tag{21}$$

where ^ stands for the estimated variables. The estimation of the virtual disturbance will be achieved as:

$$\hat{u}_{di}(t) = -K_{oi} \tanh(P_i E_{oi}(t)), \tag{22}$$

where $E_{oi}(t) = X_i(t) - \hat{X}_i(t)$, and $K_{oi}$ is a positive definite designing matrix.

**Theorem 1:** Assuming the follower agent described by (3), the observer (21) with the estimation of the virtual disturbance (22), the diagonal positive definite matrix $K_{oi}$ that satisfies (25), and a diagonal positive definite designing matrix $K_C$ in which the $j^{th}$ diagonal element $K_{Cj} > 1$, ensures that the estimation error $E_{oi}$ is uniformly ultimately bounded and the estimation of the virtual disturbance exponentially converges to a small region depending on the designing parameters.

**Proof:** We consider the following Lyapunov function:

$$V_{oi} = E_{oi}^T P_i E_{oi} \tag{23}$$

Using the agent system description (3), the observer (21), and the estimation of the virtual disturbance (22), the first time derivative of the Lyapunov function (23) is obtained as:

$$\dot{V}_{oi} = E_{oi}^T \left(A_i^T P_i + P_i A_i - P_i \sum_{\substack{j=0 \\ j\in\aleph}}^{N} \alpha_{ij} B_i B_i^T P_i\right) E_{oi} + \\
2E_{oi}^T P_i \left(\frac{1}{2}\sum_{\substack{j=0 \\ j\in\aleph}}^{N} \alpha_{ij} B_i B_i^T P_i * diag(\tanh(E_i^T P_i)) * \\
(E_{oj} - K_C \bar{E}_{oj}) - u_{di} - K_{oi} \tanh(P_i E_{oi}) + \\
B_{fi}\eta_i^T \left(\varphi_i(X_i) - \varphi_i(\hat{X}_i)\right)\right), \tag{24}$$

where $\bar{E}_{oj}$ is a column vector of $e_{oj_l} \tanh(\varrho e_{oj_l})$ for $j \in \aleph$, and $l = 1, \ldots, n$, in which $e_{oj_l}(t) = x_{j_l}(t) - \hat{x}_{j_l}(t)$ and $\varrho > 1$ is a constant. If the observer gain matrix satisfies

$$\min_n |K_{oi_n}| > \xi(\sigma_{di}^* + 2\eta_i^*) \tag{25}$$

where $\xi > 1$, and using ARE (17), then (24) can be rewritten as the following equation, when $\|P_i E_{oi}\| < \tanh^{-1}\left(\frac{1}{\xi}\right)$:

$$\dot{V}_{oi} < E_i^T(-\psi_i I_n)E_{oi} + 2E_{oi}^T P_i(\sigma_{di}^* + 2\eta_i^*)(1 - \xi \tanh(P_i E_{oi})). \tag{26}$$

So, we have

$$\dot{V}_{oi} < \begin{cases} E_i^T(-\psi_i I_n)E_{oi}; & \|P_i E_{oi}\| \geq \tanh^{-1}\left(\frac{1}{\xi}\right), \\ E_i^T(-\psi_i I_n)E_{oi} + d_{oi}; & \|P_i E_{oi}\| < \tanh^{-1}\left(\frac{1}{\xi}\right), \end{cases} \tag{27}$$

where $d_{oi} = 2(\sigma_{di}^* + 2\eta_i^*)(1 + \xi) \tanh^{-1}\left(\frac{1}{\xi}\right)$. Therefore, integrating (27) over $[0, t]$ results in (28),

$$V_{oi}(t) < \begin{cases} V_i(0)e^{-\psi_i t}; & \|P_i E_{oi}\| \geq \tanh^{-1}\left(\frac{1}{\xi}\right), \\ V_i(0)e^{-\psi_i t} + \frac{1}{\psi_i}[1 - e^{-\psi_i t}]d_{oi}; & \|P_i E_{oi}\| < \tanh^{-1}\left(\frac{1}{\xi}\right), \end{cases} \tag{28}$$

which means that Lyapunov function (23) exponentially converges to a small neighborhood of origin with the bound of $\frac{d_{oi}}{\psi_i}$, which implies that $E_{oi}$ is uniformly ultimately bounded. Furthermore, it can be inferred that the larger the value of $\xi$, the smaller region bound will be achieved. This completes the proof. ∎

Based on theorem 1, the following assumption is plausible:

$$\|B_i B_i^T \hat{u}_{di} - u_{di}\| \leq u_{di}^*, \tag{29}$$

where $u_{di}^*$ is a positive constant. The distributed neural-variable-structure adaptive control of the $i^{th}$ agent is presented in Fig. 1, in which Online Parameter Estimation (OPE) block generates the adaptive parameter $\hat{\eta}_i$. Moreover, it is illustrated that all the agents in the network are sending their information ($X_j$, for

$j \in \aleph$) to the corresponding controllers. In this subsection it is supposed that the network is not subjected to DoS attacks. The stability analysis and the proof of convergence are provided through Theorem 2.

**Theorem 2:** The neural-variable-structure controller (16) with the exogenous control vector (19) alongside the adaptive law (20) and the estimation of virtual disturbance (22), stabilizes a multi-agent system with $N$ followers and a leader described by (3)-(5) which satisfies assumptions (2) and (10). Furthermore, the control procedure, proposed by (16)-(20) and (22), ensures that the closed loop signals are asymptotically stable and all the states of the followers track the desired profile induced by the leader agent.

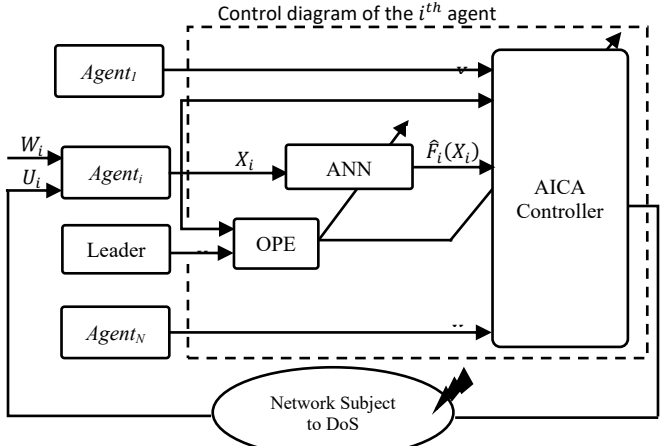

Fig. 1. Neural-variable-structure cooperative adaptive control diagram.

**Proof:** First, we consider the following Lyapunov-Krasovskii functional corresponding to the $i^{th}$ agent:

$$V_i = E_i^T P_i E_i + \tilde{\eta}_i^T K_\eta \tilde{\eta}_i \tag{30}$$

where $\tilde{\eta}_i = \hat{\eta}_i - \eta_i$, $E_i$ is the error vector (15), and $P_i = P_i^T > 0$ is obtained via ARE (17). Substituting the control input signal (16), and system descriptions (3)-(5) for the $i^{th}$ agent and the leader in the first time derivative of Lyapunov-Krasovskii functional (30) results in

$$\begin{aligned}\dot{V}_i = E_i^T(P_i A_i + A_i^T P_i)E_i + 2E_i^T P_i \Big\{(A_i - A_0)X_0 + \\ iA_i D_0 + B_{fi}f_i(X_i(t)) - B_{f0}f_0(X_0(t)) + \\ B_{\omega i}\omega_i(t) - \tfrac{1}{2}\sum_{\substack{j=0 \\ j\in\aleph}}^N \alpha_{ij}B_i B_i^T P_i \Big(E_i(t) - \\ diag(tanh(E_i^T P_i)) * E_j(t)\Big) - \\ \tfrac{1}{2}\sum_{\substack{j=0 \\ j\in\aleph}}^N \alpha_{ij} B_i B_i^T P_i K_C * diag(tanh(E_i^T P_i)) * \bar{E}_j(t) + \\ B_i B_i^T \hat{u}_{di}(t) + B_i B_i^T [B_i B_i^T + \varepsilon]^{-1}\bar{u}_{ia}\Big\} + 2\tilde{\eta}_i^T K_\eta \dot{\hat{\eta}}_i.\end{aligned} \tag{31}$$

Therefore, using $B_i B_i^T[B_i B_i^T + \varepsilon]^{-1} = I - \varepsilon[B_i B_i^T + \varepsilon]^{-1}$, and replacing the exogenous control vector (19), the estimated virtual disturbance (22) and the adaptive law (20) into (31) we have

$$\begin{aligned}\dot{V}_i = E_i^T\left(P_i A_i + A_i^T P_i - P_i \sum_{\substack{j=0\\j\in\aleph}}^N \alpha_{ij}B_i B_i^T P_i\right)E_i + \\ 2E_i^T P_i \Big\{B_{\omega i}\omega_i(t) - u_{di} + B_i B_i^T \hat{u}_{di} - \\ K_u \tanh(P_i E_i) + \tfrac{1}{2}\sum_{\substack{j=0\\j\in\aleph}}^N \alpha_{ij}B_i B_i^T P_i * \\ diag(tanh(E_i^T P_i)) * \left(E_j - K_C \bar{E}_j\right)\Big\}.\end{aligned} \tag{32}$$

As a result, because each component of diagonal matrix $K_C$ is designed to satisfy $K_{Cj} > 1$, then substituting ARE (17) into (32) and considering assumption (29) result in the following inequality if the control gain matrix $K_u$ is designed to satisfy $\|K_u\| \geq \|B_{\omega i}\|\omega_i^* + u_{di}^*$.

$$\dot{V}_i \leq E_i^T(-\psi_i I_n)E_i, \tag{33}$$

which means that regardless of the value of $\varepsilon_i^*$, the system is stable through the designed control procedure. Consequently, $V_i$, $\dot{V}_i$, and the included signals, i.e. $E_i$, $\tilde{\mu}_i$, $\dot{E}_i$, and $\dot{\hat{\eta}}_i$ are bounded. Therefore, based on Barbalat's lemma $E_i$ and $\tilde{\eta}_i$ converge to zero asymptotically, which is presented as follows:

$$\begin{aligned}V_i\big(E_i(t), \tilde{\eta}_i(t)\big) \in L_\infty &\Rightarrow \begin{cases}E_i(t) \in L_\infty, L_2, \\ \tilde{\eta}_i(t) \in L_\infty, L_2,\end{cases}\\ V_i\big(E_i(t), \tilde{\eta}_i(t)\big) \in L_\infty &\Rightarrow \dot{E}_i(t), \dot{\hat{\eta}}_i \in L_\infty \\ &\Rightarrow \begin{cases}\lim_{t\to\infty} E_i(t) = 0, \\ \lim_{t\to\infty} \tilde{\eta}_i(t) = 0.\end{cases}\end{aligned} \tag{34}$$

Therefore, based on the proposed procedure provided by (16)-(20) in the absence of DoS attacks, the agents are following the leader's profile asymptotically in spite of existence of external disturbances and unknown nonlinear function. This completes the proof. ∎

Now, let $K_\eta = P_i^{-1}(\psi_i I_n)$. Then, using (33) we have

$$\dot{V}_i \leq E_i^T(-P_i K_\eta)E_i \leq -\|K_\eta\|V_i + \tilde{\eta}_i^T K_\eta^T K_\eta \tilde{\eta}_i, \tag{35}$$

Then, using assumption (10) and (35) we have

$$\dot{V}_i(t) \leq -\varrho_i V_i(t) + d_i, \tag{36}$$

in which $\varrho_i = \|K_\eta\|$, $d_i = \|K_\eta\|^2 \varepsilon_i^{*2}$. Consequently, integrating (36) over $[0, t]$ results in

$$V_i(t) \leq V_i(0)e^{-\varrho_i t} + \tfrac{1}{\varrho_i}[1 - e^{-\varrho_i t}]d_i. \tag{37}$$

Therefore, Lyapunov function (30) exponentially converges to a small neighborhood of origin with the bound of $\tfrac{d_i}{\varrho_i} = \|K_\eta\|\varepsilon_i^{*2}$.

### 3.2 Reliability analysis in the presence of DoS attacks

During DoS attack, an adversary tries to disrupt the communication between an agent and the corresponding controller by blocking the connection. We define the set $\{t_k\}_{k \in N}$ as the sequence of time instants at which the closed-loop system begins to operate without fresh control updates for the $k^{th}$ time. In other words, $t_k$ is the last time instant at which the controller successfully transmits a new input. After this moment, the controller signal is not accessible to the agent.

In this paper, the latest successfully updated control input is retained and applied throughout the DoS interval, a strategy referred to as the Hold-Input Mechanism (HIM). This approach is in contrast to the Zero-Input Mechanism (ZIM) commonly assumed in several resilient-control frameworks, where the control input is set to zero whenever communication is interrupted. HIM offers significant advantages over ZIM: it preserves the stabilizing effect of the most recent control action, avoids abrupt transitions to open-loop dynamics, improves disturbance rejection, and prevents large tracking deviations that may lead to instability or collisions in multi-agent systems [10], [13], [28]. Moreover, HIM enables a tractable stability analysis, since the difference between the actual and last updated input remains bounded during DoS intervals, whereas ZIM may result in unbounded deviations. Therefore, in the presence of DoS attacks in the controller links ($t \in \Xi(0, +\infty)$), corresponding to the case where $\epsilon_i = 1$ in (11), the proposed control architecture applies the last available control input to the affected agent, as follows:

$$u_{ip}(t) = u_{ia}(k) = \frac{-1}{2} \sum_{\substack{j=0 \\ j \in \aleph}}^{N} \alpha_{ij} B_i^T P_i \left( E_i(k) - diag(tanh(E_i^T(k)P_i)) * E_j(k) \right) - \frac{1}{2} \sum_{\substack{j=0 \\ j \in \aleph}}^{N} \alpha_{ij} B_i^T P_i K_C diag(tanh(E_i^T(k)P_i)) \bar{E}_j(k) + B_i^T \hat{u}_{di}(k) + B_i^T [B_i B_i^T + \varepsilon]^{-1} \bar{u}_i(k), \quad (38)$$

where $u_{ip}(t)$ is the control input in the presence of DoS, $u_{ia}(k)$ is the value of control input in the absence of DoS, $u_{ia}(t)$, at time instant $t_k$, $\varepsilon$ is a positive definite diagonal matrix, $\bar{E}_j(k)$ is a column vector of $e_{jl}(k) \tanh\left(\varrho e_{jl}(k)\right)$, and $e_{jl}(k)$ is the value of $e_{jl}(t)$ at time instant $t_k$, for $j \in \aleph$, and $l = 1,2,3$, in which $\varrho > 1$ is a constant, and $K_C$ is a diagonal positive definite designing matrix in which the $j^{th}$ diagonal element $K_{Cj} > 1$. Furthermore, $P_i = P_i^T > 0$ is achieved through ARE (17). Furthermore, $\bar{u}_{ia}(k)$ and $\hat{u}_{di}(k)$ is obtained as follows:

$$\bar{u}_{ia}(k) = -B_{fi} \hat{f}_i(X_i(k)) + B_{f0} f_0(X_0(k)) - K_u \tanh(P_i E_i(k)) - (A_i - A_0) X_0(k) - iA_i D_0, \quad (39)$$

$$\hat{u}_{di}(k) = -K_{oi} \tanh(P_i E_{oi}(k)), \quad (40)$$

in which $\hat{f}_i(X_i(k))$, $E_i(k)$, and $E_{oi}(k)$ are the value of $\hat{f}_i(X_i(t))$, $E_{oi}(t)$, and $E_i(t)$ at time instant $t_k$, $D_0$, and $K_u$ are introduced in the previous section. During this time, since the neural network gain vector is assumed to be constant as $\hat{\eta}_i(k) = \hat{\eta}_i(t)|_{t=t_k}$, the adaptive law is considered as follows:

$$\dot{\hat{\eta}}_i(t) = 0, \text{ for } t \in \Xi(0, +\infty). \quad (41)$$

**Assumption 4:** As the leader's system description is considered to be known, the following assumption is feasible:

$$\|f_0(X_0(k)) - f_0(X_0(t))\| \leq \varepsilon_0. \quad (42)$$

To investigate the DoS specification which can be tolerated by the control procedure, we consider the difference between the error vector $E_i(t)$ at the current time and the value of the error vector at the last successful control input update $E_i(k)$ as follows:

$$\Pi_i(t) = E_i(t) - E_i(k). \quad (43)$$

### 3.2.1 DoS Specifications

In this subsection, we will establish the characteristic of DoS attack which is tolerable by the proposed procedure. To this end, the following lemma is proposed:

**Lemma1:** For the closed loop system (3) with the control input (38) and (39), with $\gamma_{1i} \in \mathbb{R}_{>0}$ as an appropriately designed parameter, if DoS interval $\Delta_{DoS_{ik}} = t_k^r - t_k$, subjecting the $i^{th}$ agent, satisfies

$$\Delta_{DoS_{ik}} \leq \frac{1}{\|A_i\|} \ln[1 + \gamma_{1i} \|A_i\|], \quad (44)$$

where $t_k^r$ is the time instant at which the DoS is deactivated, then

$$\|\Pi_i(t)\| \leq \gamma_{1i} \delta_{1k}, \quad (45)$$

is guaranteed, where

$$\delta_{1k} = \|B_{\omega i}\| \omega_i^* + A_i E_i(k) + \frac{1}{2} \sum_{\substack{j=0 \\ j \in \aleph}}^{N} \alpha_{ij} B_i B_i^T P_i \|E_i(k)\| + \frac{1}{2} \sum_{\substack{j=0 \\ j \in \aleph}}^{N} \alpha_{ij} B_i B_i^T P_i \|diag(tanh(P_i E_i(k)))\| \|1 - K_C \| \|E_j(k)\| + \|K_u\| \|tanh(P_i E_i(k))\| + \|B_i B_i^T\| \|\hat{u}_{di}(k)\| + \sigma_{di}^*(k) + \|B_{fi}\| \varepsilon_i^* + 2\|\hat{\mu}_i(k)\| + \|B_{f0}\| \varepsilon_0. \quad (46)$$

**Proof:** Substituting control input (38) and exogeneous control input (39) into the time derivative of (43) results in

$$\dot{\Pi}_i(t) = \dot{E}_i(t) = A_i E_i(t) + B_{\omega i} \omega_i(t) - \frac{1}{2} \sum_{j \in \aleph} \alpha_{ij} B_i B_i^T P_i \left( E_i(k) - diag(tanh(P_i E_i(k))) * E_j(k) \right) - \frac{1}{2} \sum_{j \in \aleph} \alpha_{ij} B_i B_i^T P_i K_C * diag(tanh(P_i E_i(k))) \bar{E}_j(k) + B_i B_i^T \hat{u}_{di}(k) - u_{di}(k) - K_u \tanh(P_i E_i(k)) + B_{fi} \left( f_i(X_i(t)) - \hat{f}_i(X_i(k)) \right) + B_{f0} \left( f_0(X_0(k)) - f_0(X_0(t)) \right). \quad (47)$$

On the other hand, using (6) and (9) we have

$$f_i(X_i(t)) - \hat{f}_i(X_i(k)) = \eta_i \varphi_i(X_i(t)) - \hat{\eta}_i(k)\varphi_i(X_i(k)) = \eta_i \varphi_i(X_i(t)) - \hat{\mu}_i(k)[\varphi_i(X_i(k)) + \varphi_i(X_i(t)) - \varphi_i(X_i(t))] = [\eta_i - \hat{\eta}_i(k)]\varphi_i(X_i(t)) - \hat{\eta}_i(k)[\varphi_i(X_i(k)) - \varphi_i(X_i(t))]. \quad (48)$$

Based on the activation functions (7) and assumption (10), norm of (48) can be written as

$$\|f_i(X_i(t)) - \hat{f}_i(X_i(k))\| \le \varepsilon_i^* + 2\|\hat{\eta}_i(k)\|. \quad (49)$$

Using (42), (47) and (49), we have:

$$\|\dot{\Pi}_i(t)\| \le \|A_i\|\|\Pi_i(t)\| + A_i E_i(k) + \|B_{\omega i}\|\omega_i^* + \frac{1}{2}\sum_{j \in \aleph} \alpha_{ij} B_i B_i^T P_i \|E_i(k)\| + \frac{1}{2}\sum_{j \in \aleph} \alpha_{ij} B_i B_i^T P_i \|diag(tanh(P_i E_i(k)))\| \|1 - K_C|E_j(k) + \|K_u\|\|tanh(P_i E_i(k))\| + \|B_i B_i^T\|\|\hat{u}_{di}(k)\| + \sigma_{di}^*(k) + \|B_{fi}\|\varepsilon_i^* + 2\|B_{fi}\|\|\hat{\eta}_i(k)\| + \|B_{f0}\|\varepsilon_0. \quad (50)$$

As we have $\frac{d\|\Pi_i(t)\|}{dt} \le \|\dot{\Pi}_i(t)\|$, (50) can be rewritten as

$$\|\Pi_i(t)\| \le \frac{1}{\|A_i\|}\left[e^{(\|A_i\|)(t-t_k^c)} - 1\right]\delta_{1k}. \quad (51)$$

Condition (44) implies that $\frac{1}{\|A_i\|}\left[e^{(\|A_i\|)(t-t_k^c)} - 1\right] \le \gamma_{1i}$, which completes the proof. Moreover, $\gamma_{1i}$ will be determined in the next subsection. ∎

### 3.2.2 Stability analysis

Now, we consider the following Lyapunov candidate:

$$V_i(t) = E_i^T(t) P_i E_i(t). \quad (52)$$

**Theorem 3:** In a multi-agent system containing $N$ followers and a leader described by (3)-(5), with control input (38), exogenous control input (39) and the control parameter $\gamma_{1i}$ in (45) as

$$\gamma_{1i} = \frac{\zeta_{1i}}{\delta_{1k}(\|P_i \sum_{j \in \aleph} \alpha_{ij} B_i B_i^T P_i\|)}, \quad (53)$$

and $a_1, a_2, a_3, a_4$ are designed to satisfy

$$\zeta_{1i} + \|P_i K_u\| + a_1 + a_2 + a_3 + a_5 < \gamma_i, \quad (54)$$

the Lyapunov function (52) satisfies $\dot{V}_i \le -\bar{\varrho}_i V_i(t) + \bar{d}_i$, which means that $V(t)$ exponentially converges to a small region around zero, i.e. $E_i(t)$ is uniformly ultimately bounded. Furthermore, the region bound is expressed as $\frac{\bar{d}_i}{\bar{\varrho}_i}$, where $\bar{\varrho}_i = \frac{\gamma_i - \zeta_{1i} - \|P_i K_u\| - a_1 - a_2 - a_3 - a_4}{\lambda_{max}(P_i)}$ and $\bar{d}_i = \max\{0, v - \bar{\varrho}_i(\|K_\mu\|\|\tilde{\eta}\|^2 + \|\tilde{u}_{di}\|^2)\}$, where $v$ is described as follows:

$$v = \frac{1}{a_1}\left(\|P_i \sum_{j \in \aleph} \alpha_{ij} B_i B_i^T P_i\|^2 \left\| -E_i(k) + diag(tanh(P_i E_i(k))) * (E_j(k) - K_C \bar{E}_j(k))\right\|^2\right) + \frac{1}{a_2}(\|P_i B_i B_i^T\|^2 \|\hat{u}_{di}(k)\|^2 + \sigma_{di}^{*2} \|P_i\|^2) + \frac{1}{a_3}\|P_i B_{fi}\|^2 (2\|\hat{\eta}_i(k)\| + \varepsilon_i^*)^2 + \frac{1}{a_4}(\|P_i B_{\omega i}\|\omega_i^* + \|P_i B_{f0}\|\varepsilon_0)^2. \quad (55)$$

**Proof:** Using system description (3), and plugging the control input (38), and the exogeneous control input (39) into the time derivative of (52), yield

$$\dot{V}_i = E_i^T(t)(P_i A_i + A_i^T P_i - \sum_{j \in \aleph} P_i \alpha_{ij} B_i B_i^T P_i) E_i(t) + 2E_i^T(t) P_i \{B_{fi}(\eta_i \varphi_i(X_i(t)) - \hat{\eta}_i(k)\varphi_i(X_i(k))) - B_{f0}(f_0(X_0(k)) - f_0(X_0(t))) + B_{\omega i}\omega_i(t) - \frac{1}{2}\sum_{j \in \aleph} \alpha_{ij} B_i B_i^T P_i(-E_i(t) + E_i(k) - diag(tanh(P_i E_i(k))) * E_j(k)) - \frac{1}{2}\sum_{j \in \aleph} \alpha_{ij} B_i B_i^T P_i K_C * diag(tanh(P_i E_i(k)))\bar{E}_j(k) + B_i B_i^T \hat{u}_{di}(k) - u_{di}(k) - K_u tanh(P_i E_i(k))\}, \quad (56)$$

Using (43), (49) and ARE (17), we have

$$\dot{V}_i \le E_i^T(t)(-\gamma_i I_n) E_i(t) + 2\gamma_{1i}\delta_{1k}\|P_i B_{fi}\|(\|\varepsilon_i^*\| + 2\|\hat{\eta}_i(k)\|) + 2\gamma_{1i}\delta_{1k}\varepsilon_0\|P_i B_{f0}\| + \gamma_{1i}\delta_{1k}\|\sum_{j \in \aleph} P_i \alpha_{ij} B_i B_i^T P_i\|(\|E_i(k)\| + |1 - K_C|\|E_j(k)\| + \|B_i B_i^T \hat{u}_{di}(k)\| + \|u_{di}(k)\| + \|K_u\|), \quad (57)$$

which can be rewritten as follows, based on Young's inequality on the cross terms

$$\dot{V}_i \le \left(-\gamma_i + \gamma_{1i}\delta_{1k}\|P_i \sum_{j \in \aleph} \alpha_{ij} B_i B_i^T P_i\| + \|P_i K_u tanh(P_i E_i(k))\| + a_1 + a_2 + a_3 + a_5 + a_6 + a_7\right)\|E_i(t)\|^2 + \gamma_{1i}\delta_{1k}\|P_i \sum_{j \in \aleph} \alpha_{ij} B_i B_i^T P_i\| + \|P_i B_{fi}\|\|\hat{\eta}_i(k)\|\|\varphi_i(X_i(k)) - \varphi_i(X_i(t))\|^2 + \frac{1}{a_6}(\|P_i B_{\omega i}\|\omega_i^*)^2 + \frac{1}{a_7}(\|P_i B_{f0}\|\varepsilon_0)^2 + \frac{1}{a_1}\left(\|P_i \sum_{j \in \aleph} \alpha_{ij} B_i B_i^T P_i\|^2 \left\|-E_i(k) + diag(tanh(P_i E_i(k))) * (E_j(k) - K_C \bar{E}_j(k))\right\|^2\right) + \frac{1}{a_2}(\|P_i B_i B_i^T\|^2 \|\hat{u}_{di}(k)\|^2) + \frac{1}{a_3}\sigma_{di}^{*2}\|P_i\|^2 + \frac{1}{a_5}(\|\eta_i^* - \hat{\eta}_i(k)\|^2 \|P_i B_{fi}\|^2) + \|P_i K_u tanh(P_i E_i(k))\| \le \left(-\gamma_i + \gamma_{1i}\delta_{1k}\|P_i \sum_{j \in \aleph} \alpha_{ij} B_i B_i^T P_i\| + \|P_i K_u tanh(P_i E_i(k))\| + a_1 + a_2 + a_3 + a_4 + a_5\right)\|E_i(t)\|^2 + \frac{1}{a_1}\left(\|P_i \sum_{j \in \aleph} \alpha_{ij} B_i B_i^T P_i\|^2 \left\|-E_i(k) + diag(tanh(P_i E_i(k))) * (E_j(k) - K_C \bar{E}_j(k))\right\|^2\right) + \quad (58)$$

$\frac{1}{a_2}(\|P_i B_i B_i^T\|^2 \|\hat{u}_{di}(k)\|^2 + \sigma_{di}^{*2}\|P_i\|^2) +$
$\frac{1}{a_3}(2\|P_i B_{fi}\|\|\hat{\eta}_i(k)\|)^2 + \frac{1}{a_4}(\|\eta_i^* -$
$\hat{\eta}_i(k)\|^2 \|P_i B_{fi}\|^2) + \frac{1}{a_5}(\|P_i B_{\omega i}\|\omega_i^* + \|P_i B_{f0}\|\varepsilon_0)^2.$

Substituting (53) into (58) results in

$$\dot{V}_i \leq (-\gamma_i + \zeta_{1i} + \|P_i K_u \tanh(P_i E_i(k))\| + a_1 + a_2 + a_3 + a_4 + a_5)\|E_i(k)\|^2 + \zeta_{1i} +$$
$$\|P_i B_{fi}\|\|\hat{\eta}_i(k)\|\|\varphi_i(X_i(k)) - \varphi_i(X_i(t))\|^2 +$$
$$\|P_i B_{f0}\|\|f_0(X_0(k)) - f_0(X_0(t))\|^2 +$$
$$\|P_i B_{\omega i}\|\|\omega_i(t)\|^2 +$$
$$\frac{1}{a_1}\left(\|P_i \sum_{j \in \aleph} \alpha_{ij} B_i B_i^T P_i\|^2 \| -E_i(k) + \text{diag}(\tanh(P_i E_i(k))) * (E_j(k) - K_C \bar{E}_j(k))\|^2\right) +$$
$$\frac{1}{a_2}(\|P_i B_i B_i^T\|^2 \|\hat{u}_{di}(k)\|^2 + \sigma_{di}^{*2}\|P_i\|^2) +$$
$$\frac{1}{a_3}(2\|P_i B_{fi}\|\|\hat{\eta}_i(k)\|)^2 + \frac{1}{a_4}(\|\eta_i^* - \hat{\eta}_i(k)\|^2 \|P_i B_{fi}\|^2) + \frac{1}{a_5}(\|P_i B_{\omega i}\|\omega_i^* + \|P_i B_{f0}\|\varepsilon_0)^2. \quad (59)$$

Considering (54) in (59) we have $\dot{V}_i \leq -\bar{\varrho}_i V_i(t) + \bar{d}_i$, which results in

$$V_i(t) \leq e^{-\bar{\varrho}_i(t - t_i^l)} V_i(t_i^l) + \frac{1}{\bar{\varrho}_i}\left[1 - e^{-\bar{\varrho}_i(t - t_i^l)}\right] \bar{d}_i, \quad (60)$$

where $\bar{\varrho}_i$ and $\bar{d}_i$ are introduced in Theorem 3. This completes the proof. ■

### 3.3 Reliability analysis of the closed loop system under DoS attack

We consider $\Lambda(\tau, t)$ as the union of sub-intervals of $[\tau, t)$ at which there is no DoS and the multi-agent system, with $N$ followers (3) and a leader (4), is operating under control. On the other hand, $\Xi(\tau, t)$ is assumed to be sub-intervals of $[\tau, t)$ at which DoS attack is active in controller links and the subjected agents are operating out of control. The secure adaptive artificial-intelligence-based control is proposed based on the following switching method:

$$u_i(t) = \begin{cases} u_{ia}(t) & \text{if } t \in \Lambda(\tau, t); \\ u_{ip}(t) = u_{ia}(k) & \text{if } t \in \Xi(\tau, t), \end{cases} \quad (61)$$

where $u_{ia}(t)$ is the control input in the absence of DoS attacks proposed by (16); and $u_{ip}(t)$ is the control input vectors in the presence of DoS attacks on the controller links, provided by (38) with control parameters (53) and (54). Moreover, the exogenous control vector is proposed as follows:

$$\bar{u}_i(t) = \begin{cases} \bar{u}_{ia}(t) & \text{if } t \in \Lambda(\tau, t); \\ \bar{u}_{ip}(t) = \bar{u}_{ia}(k) & \text{if } t \in \Xi(\tau, t), \end{cases} \quad (62)$$

in which $\bar{u}_{ia}(t)$, and $\bar{u}_{ip}(t)$ stand for the exogenous control vector proposed for the absence of DoS attacks (19), and the presence of DoS attacks on the controller links (39), respectively. Furthermore, $u_{ia}(k)$ and $\bar{u}_{ia}(k)$, in (61) and (62), are respectively the values of $u_{ia}(t)$ and $\bar{u}_{ia}(t)$ at time instant $t_k$. In addition, the estimation of virtual disturbance, based on (22) and (40), and the adaptive law for neural network gain vector, using (20) and (41), are as follows:

$$\hat{u}_{di}(t) = \begin{cases} -K_{oi} \tanh(P_i E_{oi}(t)) & \text{if } t \in \Lambda(\tau, t); \\ -K_{oi} \tanh(P_i E_{oi}(k)) & \text{if } t \in \Xi(\tau, t); \end{cases} \quad (63)$$

$$\dot{\hat{\mu}}_i(t) = \begin{cases} K_\mu^{-1} \varphi_i(x_i(t)) E_i^T(t) P_i B_{fi} & \text{if } t \in \Lambda(\tau, t); \\ 0 & \text{if } t \in \Xi(\tau, t); \end{cases} \quad (64)$$

For the sake of simplicity, the proposed procedure is provided by algorithm 1.

---

**Algorithm 1:** Distributed Cyber-secure adaptive control at agent $i$ under DoS attacks:

**Input:** $x_i(t)$, $u_i(t)$ and $x_j(t)$, $\forall j = 0, \dots, N$

**Start:** for $j = 0$ to $N$ do

  **if** $u_i(t) \neq 0$ and $\dot{X}_0(t) \neq 0$, $\forall j = 0, \dots, N$ **then**

    There is no DoS attack in the system;

    Use controller (16) with exogenous control vector (19), the estimation of virtual disturbance (22) and adaptive law (20).

  **else if** $u_i(t) = 0$ and $\dot{X}_0(t) \neq 0$ **then**

    The $i^{th}$ controller link is subjected to DoS attacks;

    Use controller (38) with exogenous control vector (39), the estimation of virtual disturbance (40) and adaptive law (41).

end

---

**Theorem 4:** A multi-agent system containing $N$ followers and a leader described by (3)-(5) under DoS attacks satisfying assumptions 1 and 2, with switching control input (61), exogenous control input (62), the estimation of virtual disturbance (63), and adaptive law (64) is stable. Furthermore, the Lyapunov function $V(t)$ exponentially converges to a small neighborhood of origin with the region bound of $\left[1 + \frac{2e^{\chi_0(1-\frac{1}{T})n_0\tau_D}}{1 - e^{-\chi_0(1-\frac{1}{T})\tau_D}}\right]\rho^*$, where $\rho^* = max\left(\frac{v_0}{\chi_0}, \frac{v_1}{\chi_1}\right)$.

**Proof:** We denote that $\|K_\mu\| = \chi_0$, $d_i = v_0$, $\bar{\varrho}_i = \chi_1$ and $\bar{d}_i = v_1$. So, the time derivative of the Lyapunov function can be written as follows:

$$\dot{V}_i(t) \leq -\chi V_i(t) + v, \quad (65)$$

where $\chi = \chi_0$ and $v = v_0$ for $t \in \Lambda(0, +\infty)$ and $\chi = \chi_1$ and $v = v_1$ for $t \in \Xi(0, +\infty)$. Therefore, for $t \in [t_\sigma, t_{\sigma+1})$ we use $\chi = \chi_j$ to present one of on/off transitions of DoS attacks in the controller links. So, we have

$$V_i(t) \leq V_i(t_\sigma)e^{-\chi_\sigma(t-t_\sigma)} \tag{66}$$

Using the recursive solution for $[0,t]$, we have

$$V_i(t) \leq V_i(0)e^{-(\chi_0|\Lambda(0,t)|+\chi_1|\Xi(0,t)|)} + \left[1 + \sum_{t_\sigma \leq t, \sigma \in \mathbb{N}_+} 2e^{-\chi_0|\Lambda(t_\sigma,t)|-\chi_1|\Xi(t_\sigma,t)|}\right]\rho^*, \tag{67}$$

where $\rho^* = max\left(\frac{v_0}{\chi_0}, \frac{v_1}{\chi_1}\right)$. According to assumption 2 on the energy of the adversary, we have

$$0 \leq |\Xi(\tau,t)| \leq \frac{t-\tau}{T}, \tag{68}$$

$$|\Lambda(\tau,t)| \geq t - \tau - |\Xi(\tau,t)| \geq \left(1 - \frac{1}{T}\right)(t-\tau), \tag{69}$$

Therefore, we have

$$-\chi_1\left(\frac{t-\tau}{T}\right) \leq -\chi_1|\Xi(\tau,t)| \leq 0, \tag{70}$$

$$-\chi_0|\Lambda(\tau,t)| \leq -\chi_0\left(1 - \frac{1}{T}\right)(t-\tau), \tag{71}$$

Substituting (70) and (71) into (67) results in

$$V_i(t) \leq V_i(0)e^{-\chi_0\left(1-\frac{1}{T}\right)t} + \left[1 + 2\sum_{t_\sigma \leq t, \sigma \in \mathbb{N}_+} e^{-\chi_0\left(1-\frac{1}{T}\right)(t-t_\sigma)}\right]\rho^*, \tag{72}$$

Using assumption 1 on DoS frequency, we have $t - t_\sigma \geq (n(t_\sigma, t) - n_0)\tau_D$. Therefore, (72) can be rewritten as

$$V_i(t) \leq V_i(0)e^{-\chi_0\left(1-\frac{1}{T}\right)t} + \left[1 + 2\sum_{t_\sigma \leq t, \sigma \in \mathbb{N}_+} e^{-\chi_0\left(1-\frac{1}{T}\right)(n(t_\sigma,t)-n_0)\tau_D}\right]\rho^*. \tag{73}$$

Using the definition of $t_k$, $k$ denotes the number of times in which a DoS attack is activated in the controller links before the current time t. Therefore, $n(t_\sigma, t) \geq k - \sigma$, for $\sigma \leq k$. As a result, we have

$$V_i(t) \leq V_i(0)e^{-\chi_0\left(1-\frac{1}{T}\right)t} + \left[1 + 2e^{\chi_0\left(1-\frac{1}{T}\right)n_0\tau_D}\sum_{\sigma=1}^{j}e^{-\chi_0\left(1-\frac{1}{T}\right)(j-\sigma)\tau_D}\right]\rho^*. \tag{74}$$

Summing the sequence $\left\{e^{-\chi_0\left(1-\frac{1}{T}\right)(j-\sigma)\tau_D}\right\}_{\sigma \leq j, \sigma \in \mathbb{N}_+}$ results in:

$$\begin{aligned}V_i(t) &\leq V_i(0)e^{-\chi_0\left(1-\frac{1}{T}\right)t} \\ &+ \left[1 + 2e^{\chi_0\left(1-\frac{1}{T}\right)n_0\tau_D}\left(\frac{1-e^{-j\chi_0\left(1-\frac{1}{T}\right)\tau_D}}{1-e^{-\chi_0\left(1-\frac{1}{T}\right)\tau_D}}\right)\right]\rho^* \\ &\leq V_i(0)e^{-\chi_0\left(1-\frac{1}{T}\right)t} + \left[1 + \frac{2e^{\chi_0\left(1-\frac{1}{T}\right)n_0\tau_D}}{1-e^{-\chi_0\left(1-\frac{1}{T}\right)\tau_D}}\right]\rho^*.\end{aligned} \tag{75}$$

Therefore, Lyapunov function exponentially converges to a residual bound $\left[1 + \frac{2e^{\chi_0\left(1-\frac{1}{T}\right)n_0\tau_D}}{1-e^{-\chi_0\left(1-\frac{1}{T}\right)\tau_D}}\right]\rho^*$, which implies that $E_i$, $\tilde{\mu}_i$ and $\tilde{u}_{di}$ are uniformly ultimately bounded under DoS attack. This completes the proof. ∎

Moreover, it is shown through theorem 2, and theorem 3 that when there is no attack in the system, $E_i$ and $\tilde{\mu}_i$ asymptotically converge to zero, and if there exists DoS attacks in the controller links, $E_i$ and $\tilde{\mu}_i$ are uniformly ultimately bounded.

## 4 SIMULATION RESULTS AND ANALYSIS

In this section, the performance of the proposed neural-variable-structure procedure is scrutinized via simulation studies through Matlab/Simulink as an adaptive resilient cruise control (ARCC) in a platoon of $N$ connected automated vehicles (CAVs) and a leader.

### 4.1 Network topology and vehicular model

We assume of 5 CAVs and a leader, travelling in a platoon on a strait road, with the following longitudinal dynamics for the $i^{th}$ follower [8], [29], [30]:

$$\begin{aligned}\dot{p}_i(t) &= v_i(t), \\ \dot{v}_i(t) &= a_i(t), \\ \dot{a}_i(t) &= \frac{-1}{\tau_i}a_i(t) + \frac{1}{m_i\tau_i}u_i(t) + f_i(x_i(t)) + \omega_i(t),\end{aligned} \tag{76}$$

where $p_i$, $v_i$, $a_i$ and $m_i$ are the position, the velocity, the acceleration, and the mass of the $i^{th}$ vehicle, respectively, $\tau_i$ is the drivetrain time constant, $u_i$ is the control input. Furthermore, $\omega_i$ represents the unknown external disturbances, which are caused by contextual effects such as wind velocity, and air specifications; and $f_i(x_i(t))$ denotes the unknown nonlinear function, including rolling and hill climbing resistance force, and other nonlinear terms of a vehicle, which are impractical to be determined precisely. Considering $X_i(t) = [p_i(t) \quad v_i(t) \quad a_i(t)]^T$, results in system description (3) with the following matrices

$$A_i = \begin{bmatrix} 0 & 1 & 0 \\ 0 & 0 & 1 \\ 0 & 0 & \frac{-1}{\tau_i} \end{bmatrix}, B_i = \begin{bmatrix} 0 \\ 0 \\ \frac{1}{\tau_i} \end{bmatrix}, C_i = \begin{bmatrix} 1 & 0 & 0 \\ 0 & 1 & 0 \\ 0 & 0 & 1 \end{bmatrix}, \\ B_{fi} = B_{\omega i} = \begin{bmatrix} 0 \\ 0 \\ 1 \end{bmatrix}. \tag{77}$$

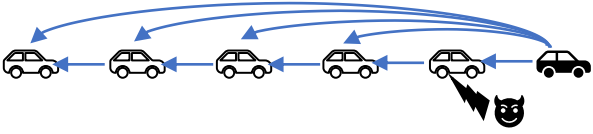

Fig. 2. LPF Network topologies of CAVs in the platoon.

Furthermore, it is supposed that the leader is exposing a tracking profile to other vehicles to follow as shown in Fig. 3. Moreover, no external disturbance or attack is supposed to affect the leader system description.

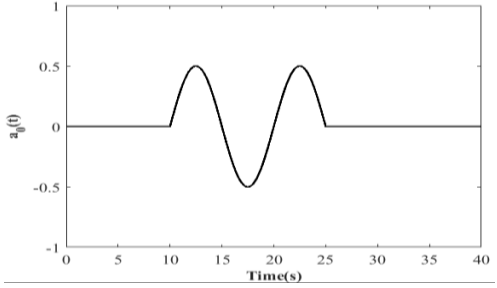

Fig. 3. Driving profile exposed by the leader.

On the other hand, leader-predecessor follower (LPF), shown in Fig. 2, in which each following agent receives the information of the leader and the predecessor vehicle through a nonreliable vehicle-to-vehicle communication network, is assumed for the network topology. Moreover, vehicle and controller parameters are provided through Table 1 and Table 2:

TABLE 1. PARAMETERS OF CAVS [31]

| $Veh_i$ / Parameter | Leader | $Veh_1$ | $Veh_2$ | $Veh_3$ | $Veh_4$ | $Veh_5$ |
|---|---|---|---|---|---|---|
| $\tau$ | 0.1 | 0.1 | 0.3 | 0.5 | 0.7 | 0.9 |
| M | 1500 | 1500 | 2000 | 2500 | 3000 | 3500 |
| $d_m$ | 5 | 5 | 7 | 9 | 11 | 13 |
| $\sigma$ | 1.2 | 1.2 | 1.5 | 1.8 | 2.1 | 2.4 |
| S | 2.2 | 2.2 | 4.2 | 6.2 | 8.2 | 10.2 |
| $C_d$ | 0.35 | 0.35 | 0.4 | 0.45 | 0.5 | 0.55 |
| $\rho$ | 30° | 30° | 30° | 30° | 30° | 30° |
| $f$ | 0.02 | 0.02 | 0.04 | 0.06 | 0.08 | 0.1 |
| $d_0$ | 7 | 7 | 7 | 7 | 7 | 7 |

TABLE 2. CONTROLLER PARAMETERS

| $\gamma_i$ | $K_{u_i}$ | $K_{o_i}$ | $K_{f_i}$ | $K_\mu$ | $\varepsilon$ |
|---|---|---|---|---|---|
| 1000 | 1000 $I_3$ | 1000 $I_3$ | 1000 $I_3$ | 1000 $I_3$ | 0.5 $I_3$ |

where matrix $I_3$ is a 3 × 3 identity matrix.

### 4.2 Neural network estimator specifications

At this stage, the optimum number of neurons in the hidden layer of ANN will be specified. To this end, the error signals at $t = 25s$, generated through different number of neurons in the hidden layer are evaluated via Fig. 4.

It can be inferred from Fig. 4 that the more neurons in the hidden layer results in the better precision in terms of position and acceleration. However, to prohibit the excessive computational burden on the control procedure, 25 neurons are assumed considered in the hidden layer, which consequences a more precise result in the velocity error signal as well as to an acceptable precision on the other two error signals.

### 4.3 Reliable Adaptive NVSC results in the absence of DoS attacks

Now, we assume that there is no DoS attack in the system to verify the proposed reliable adaptive neural-variable-structure control (NVSC) procedure on a healthy application. The observer error signals are provided via Fig. 6 through which it can be inferred that the error signals of the proposed virtual disturbance observer are uniformly ultimately bounded (UUB) and converge to a small region. For the sake of clarification, position, velocity and acceleration of each vehicle is provided in Fig. 7 through which it is shown that all the vehicles in the platoon are following the velocity and the acceleration profile of the leader (dashed line) as well as maintaining a predefined safe distance between each two consecutive vehicles, which means that no collision is happened in the fleet

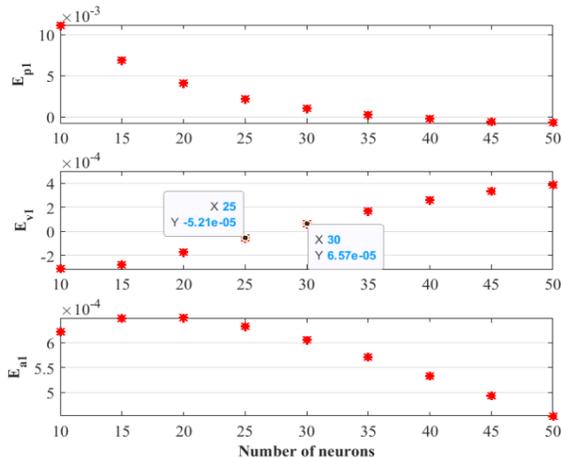

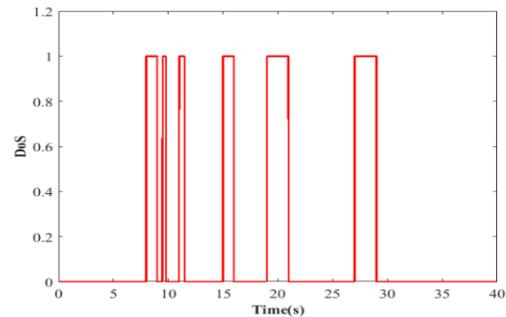

Fig. 5. DoS attack signal injected into the controller link of the first follower.

Fig. 4. Effect of number of neurons in the hidden layer on steady-state errors of the position, velocity, and acceleration..

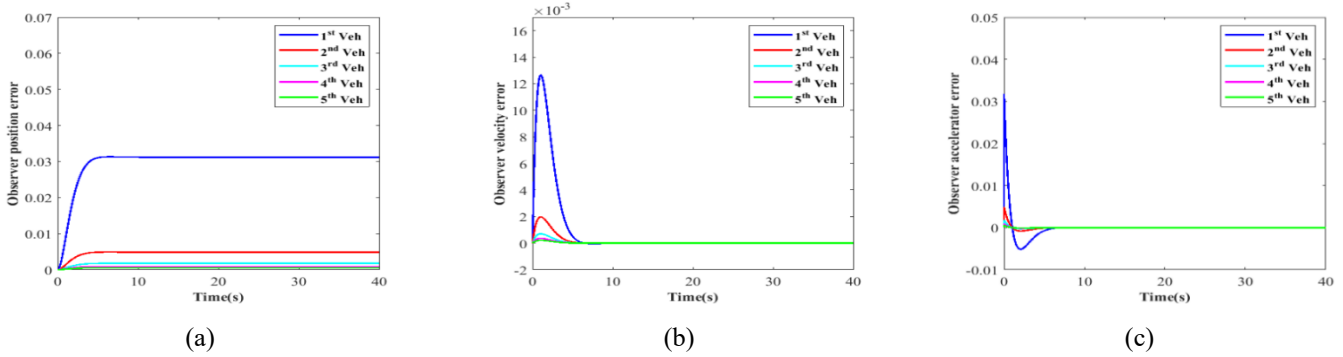

(a) (b) (c)

Fig. 6. Observer error signals of (a) Position (b) velocity, and (c) acceleration of the followers and the leader in a platoon (No DoS attack)

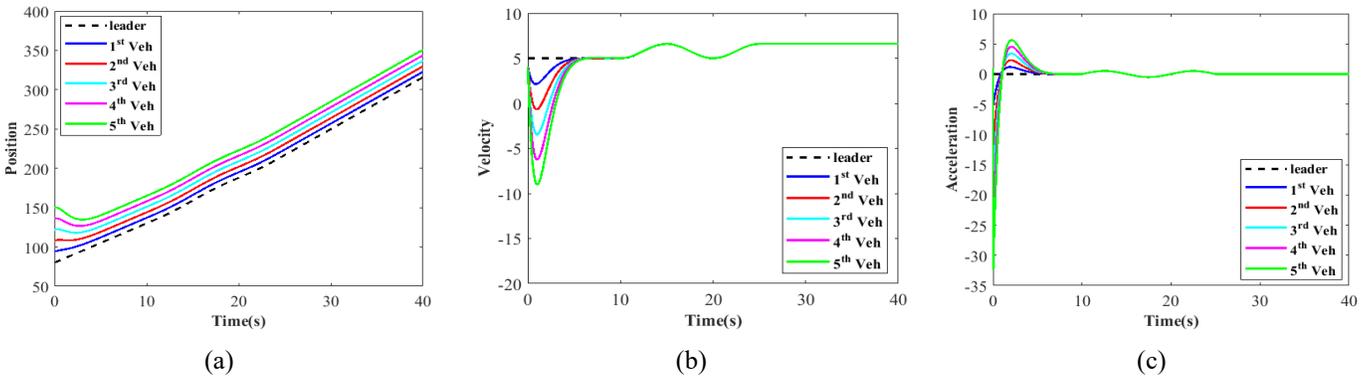

(a) (b) (c)

Fig. 7. Evolution of followers and leader's (a) position (b) velocity, and (c) acceleration in a platoon with LPF topology (No DoS attack)

### 4.4 Reliable Adaptive NVSC results in the presence of DoS attacks

To scrutinize the proposed Reliable Adaptive NVSC procedure on the system performance under DoS attacks, it is assumed the 1st vehicle in the platoon is subjected to a sustained DoS signal, which is modelled by a rectangular wave with a variable random duty cycle, which is shown in Fig. 5. Furthermore, position, velocity, and acceleration of each follower while there exist DoS attacks in the first vehicle are provided in Fig. 8.

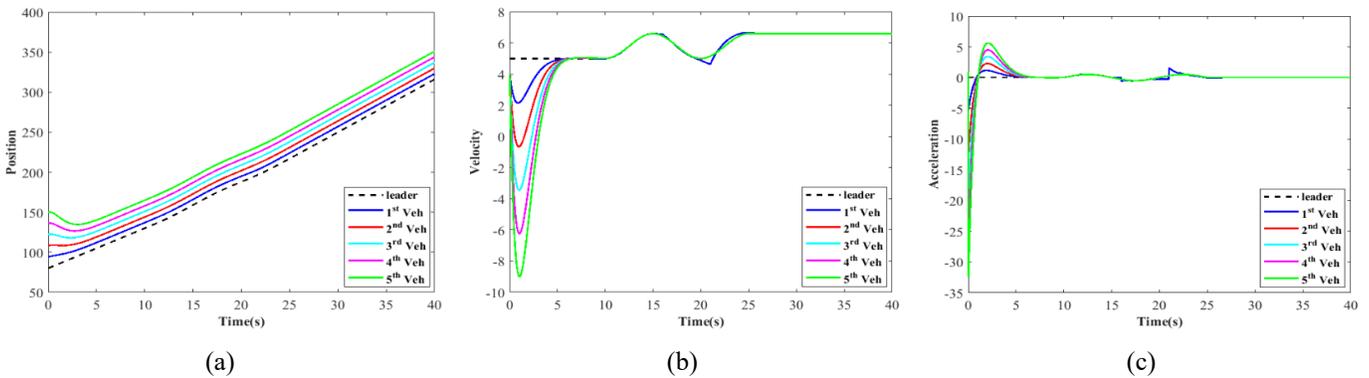

(a) (b) (c)

Fig. 8. (a) Position, (b) velocity, and (c) acceleration of the followers with respect to the leader's profile. (with DoS attack).

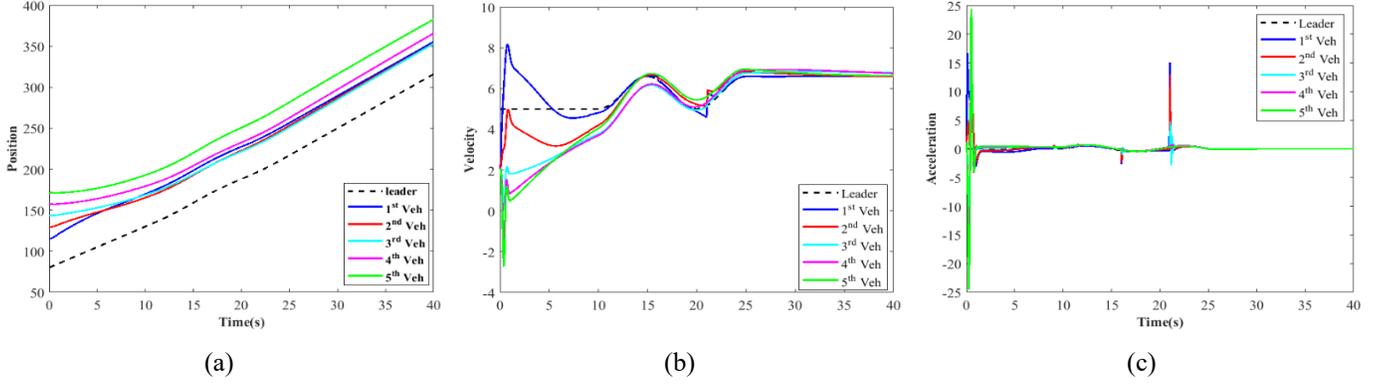

Fig. 9. (a) Error signals of the position, (b) the velocity, and (c) the acceleration of the following vehicles with respect to the leading one through the secure adaptive control procedure in [7].

It is shown through Fig. 8 that when DoS attacks occur, i.e. DoS=1, the communication of the controller to the corresponding vehicle is blocked, and the subjected agent is using the latest update of the control signal on hand presented by (38). Moreover, it is shown that although there are some deviations in the velocity and acceleration evolution due to the DoS attacks in the first agent, all the agents are following the leader's profile in terms of velocity and acceleration while maintaining the desired predefined distance between each two consecutive vehicles. Furthermore, the positions of the followers show that no collision occurred neither between any two followers nor between the leader and the first vehicle. Furthermore, the proposed method in this paper is compared to the protocol in [7], to illustrate the effectiveness of the designed procedure shown in Fig. 9

As it can be inferred from Fig. 9 (a), in the case of DoS attack in the controller link of the first vehicle, the proposed procedure in [7] is unable to control the position of the followers, as the desired safe distance is not maintained between each two consecutive vehicles and some collisions have happened in their travel. For instance: 1) between the 1st and the 2nd vehicles at $t = 5.8s$, 2) between the 1st and the 3rd vehicles at $t = 9.2s$, 3) between the 2nd and the 3rd vehicles at $t = 17s$. Moreover, it is shown in Fig. 9(b), and Fig. 9(c) that the procedure proposed in [7] is incapable to control the followers in terms of velocity and acceleration when there exists DoS attacks. They are also extremely slow to converge to the desired profile of the leader.

## 5 DISCUSSION

This work developed a reliable adaptive neural–variable-structure control framework for nonlinear multi-agent systems operating under Denial-of-Service attacks and potentially singular control gains. The simulation results presented in Section 4 provide several important insights into the behavior and practical relevance of the proposed method.

First, the results demonstrate that the combination of neural estimation, virtual disturbance observation, and variable-structure compensation leads to robust tracking performance even when the nonlinear dynamics of the agents are completely unknown. The adaptive neural network effectively approximates the nonlinear components, while the disturbance observer mitigates the influence of external perturbations and singularities in the control gain. As shown in the no-attack scenario, all followers successfully track the leader's trajectory while maintaining safe inter-vehicle distances, validating the theoretical stability guarantees.

A second key observation concerns the system behavior under DoS attacks. When the communication between a controller and its corresponding agent is intermittently obstructed, the proposed reliability rule and switched control architecture play a decisive role in maintaining stability. In particular, the adoption of a *Hold-Input Mechanism (HIM)*, where the last successfully updated control input is retained during attack intervals, prevents the abrupt loss of actuation that commonly destabilizes nonlinear or interconnected systems. The simulation with a randomly varying DoS duty cycle confirms that HIM effectively limits error growth, preserves string stability in the vehicle platoon, and avoids collisions. These results align with established findings in secure and networked control showing that sample-and-hold behavior provides significantly greater robustness than zero-input policies during communication outages.

The comparison with the existing secure adaptive control procedure in [7] further highlights these advantages. Under the same attack profile and model parameters, the method in [7] exhibits substantial deviations in position, velocity, and acceleration, ultimately resulting in multiple collisions. This degradation is primarily due to the Zero-Input Mechanism employed during DoS intervals and the absence of a reliability-based switching rule. The contrasting outcomes of Fig. 8 and Fig. 9 therefore provide strong empirical evidence that the proposed framework not only achieves theoretical resilience but also delivers meaningful safety benefits in practical cyber-physical systems such as connected automated vehicles.

It is also notable that the computational burden of the proposed controller remains manageable. Although neural networks increase the complexity of online adaptation, the number of neurons can be tuned to balance precision and computational efficiency. The analysis in Fig. 4 suggests that a moderate number of neurons provides sufficient accuracy for disturbance compensation without imposing excessive overhead, an important consideration for embedded implementation.

Nonetheless, several limitations and opportunities for future research should be acknowledged. While the simulations consider a realistic vehicle platoon model and DoS pattern, experimental validation on hardware platforms or hardware-in-the-loop environments would further substantiate the applicability of the approach. Extending the framework to partially connected networks, switching topologies, or heterogeneous communication delays also represents a promising direction. Finally, integrating predictive or learning-based DoS forecasters may further enhance resilience in adversarial settings.

Overall, the discussion above confirms that the proposed method provides a reliable, scalable, and practically meaningful solution for resilient coordination of nonlinear multi-agent systems under communication-level attacks and actuator singularities.

## 6 CONCLUSION

This paper addressed the problem of reliable control for nonlinear cyber-physical multi-agent systems subject to singular control gains and Denial-of-Service (DoS) attacks, two critical challenges that can severely degrade stability and coordination in distributed networks. A neural-variable-structure adaptive control framework was developed, incorporating a reliability-assessment rule and a switched control architecture that ensures resilience against communication interruptions. The proposed method employs neural-network-based estimation to handle unknown nonlinearities, and a virtual disturbance observer to accommodate singular control gains, thereby removing restrictive assumptions commonly found in existing resilient MAS control strategies.

Rigorous Lyapunov-based analysis established uniform ultimate boundedness of all closed-loop signals and guaranteed consensus tracking in both the absence and presence of DoS attacks. Simulation studies on a vehicle platoon further validated the effectiveness of the method, demonstrating robust performance, reduced tracking errors, and clear advantages over existing approaches.

The results confirm that the proposed framework provides a unified, scalable, and reliable solution for resilient coordination in cyber-physical systems with challenging actuator and communication constraints. Future work may extend the method to time-varying DoS patterns, partially connected networks, or hardware-in-the-loop implementations.

STATEMENTS AND DECLARATIONS:

**Funding:** This research was supported by the Natural Sciences and Engineering Research Council of Canada (NSERC), which is gratefully acknowledged.

**Conflict of Interest:** The authors declare that they have no conflict of interest.